\documentclass[aps,prb,twocolumn,superscriptaddress,showpacs]{revtex4-1}

\usepackage{graphicx}
\usepackage{calc}
\usepackage{bm}
\usepackage{color}

\begin{document}

\title{Spin-dependent transport through the Weyl semimetal surface}

\author{V.D.~Esin}
\author{D.N.~Borisenko}
\author{A.V.~Timonina}
\author{N.N.~Kolesnikov}
\author{E.V.~Deviatov}
\affiliation{Institute of Solid State Physics of the Russian Academy of Sciences, Chernogolovka, Moscow District, 2 Academician Ossipyan str., 142432 Russia}

\date{\today}

\begin{abstract}
We experimentally  compare two types of interface structures with magnetic and non-magnetic Weyl semimetals. They are the junctions between a gold normal layer and magnetic Weyl semimetal Ti$_2$MnAl, and a ferromagnetic nickel layer and non-magnetic Weyl semimetal WTe$_2$, respectively. Due to the ferromagnetic side of the junction, we investigate spin-polarized transport through the Weyl semimetal surface. For both  structures, we demonstrate similar current-voltage characteristics, with hysteresis at low currents and sharp peaks in differential resistance at high ones. Despite this behavior resembles the known current-induced magnetization dynamics in ferromagnetic structures, evolution  of the resistance peaks with magnetic field is unusual.  We connect the observed effects  with current-induced spin dynamics in Weyl topological surface states.
\end{abstract}


\maketitle

\section{Introduction}

Recent interest to topological semimetals is connected with their peculiar properties~\cite{armitage},  which originates from gapless spectrum with band touching in some distinct points. In Weyl semimetals (WSM) every touching point splits  into two Weyl nodes with opposite chiralities due to the time reversal or inversion symmetries breaking. The projections of two Weyl nodes on the surface Brillouin zone are connected by a Fermi arc, which represents the topologically protected  surface state~\cite{armitage}. Most of experimentally investigated WSMs, were non-centrosymmetric crystals with broken inversion symmetry~\cite{armitage}. For example,  spin- and angle- resolved photoemission spectroscopy data indeed demonstrate  spin-polarized surface Fermi arcs~\cite{das16,feng2016} for a WTe$_2$  Weyl semimetal~\cite{li2017,soluyanov}. In contrast, there are only a few candidates of magnetically ordered materials for the realization of WSMs~\cite{mag1,mag2,mag3,mag4,kagome,kagome1}. 

Ti$_2$MnAl is one of the newly predicted~\cite{timnal,timnal_review} magnetic WSM. The bulk Ti$_2$MnAl is a spin gapless semiconductor, where  the valence and conduction bands touch each other in the spin-up channel and there is a large gap in the spin down band structure~\cite{timnal_exp}. Therefore, the bulk Ti$_2$MnAl has 100\% spin polarized carriers.

It is well known, that the magnetically ordered materials allows complicated magnetization dynamics. For example, current-induced excitation of spin waves, or magnons,  was demonstrated as sharp $dV/dI$ differential resistance  peaks in ferromagnetic multilayers at large electrical current densities~\cite{myers,tsoi1,tsoi2,katine,single,balkashin,balashov}. In these structures, spin-dependent scattering may even reverse the magnetic moments of the  layers, which results in  $dV/dI$ switchings at low currents, accomplished by well-defined hysteresis~\cite{myers}. 

Bulk magnons were also demonstrated~\cite{cosns} for magnetic WSM at low current densities due to the coupling between two magnetic moments mediated by Weyl fermions~\cite{weyl_magnon}. Also, in a bilayer consisting of a magnetic WSM and a normal metal,  a charge current can be induced in the WSM by  spin current injection at the interface~\cite{bilayer}.  On the other hand, there are spin-polarized surface Fermi arcs on a WSM  surface~\cite{das16,feng2016,jiang15,rhodes15,wang16}. Similarly to the case of topological insulators~\cite{topinssurf}, one can expect current-induced magnetization dynamics~\cite{current} also for  surface magnetic textures~\cite{texture,araki} in WSM.

Here, we experimentally  compare two types of interface structures with magnetic and non-magnetic Weyl semimetals. They are the junctions between a gold normal layer and magnetic Weyl semimetal Ti$_2$MnAl, and a ferromagnetic nickel layer and non-magnetic Weyl semimetal WTe$_2$, respectively. Due to the ferromagnetic side of the junction, we investigate spin-polarized transport through the Weyl semimetal surface. For both  structures, we demonstrate similar current-voltage characteristics, with hysteresis at low currents and sharp peaks in differential resistance at high ones. Despite this behavior resembles the known current-induced magnetization dynamics in ferromagnetic structures, evolution  of the resistance peaks with magnetic field is unusual.  We connect the observed effects  with current-induced spin dynamics in Weyl topological surface states.

\section{Samples and technique}

\begin{figure}
\includegraphics[width=\columnwidth]{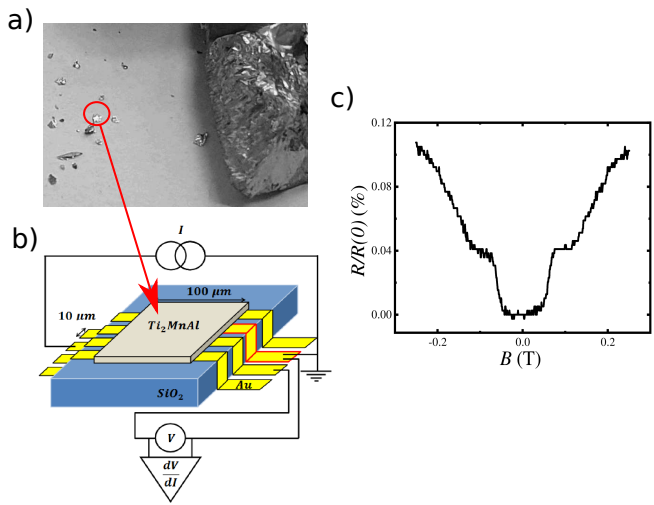}
\caption{(Color online) (a) An initial Ti$_2$MnAl drop (right) and  cleaved flakes (left). (b) The sketch of a sample with electrical connections. 100 nm thick and 10 $\mu$m wide Au leads are formed on a SiO$_2$ substrate. A Ti$_2$MnAl flake  ($\approx$ 100~$\mu$m size, denoted by red circle in (a))  is transferred on top of the leads with $\approx$ 10 $\mu$m overlap, forming planar Au-Ti$_2$MnAl  junctions. Charge transport is investigated with a standard three-point technique: the studied contact (denoted by the red border) is grounded and two other contacts are used for applying current and measuring potential. (c) The bulk  Ti$_2$MnAl material demonstrates low positive magnetiresistance which coincide even quantitatively with the known one~\cite{timnal_exp} for this material. 
}
\label{sample}
\end{figure}

Ti$_2$MnAl was obtained as a bulk ingot by levitation melting in high-frequency (60-70 kHz) induction furnace. A mixture of Mn and Al powders was placed into the cylindrical titanium capsule and melted in a suspended condition for 20 minutes in argon medium at 0.2 MPa pressure and at 2080 K temperature. After switching the heater off the resulting globule of the melt was dropped to a cooled copper crystallizer, where it was quenched at 278 K. The ingot cleaved mechanically for further processing as shown in Fig.~\ref{sample} (a) and (b). We check by standard magnetoresistance measurements that our Ti$_2$MnAl is characterized by low positive magnetiresistance, see Fig.~\ref{sample} (c), which has been demonstrated for this material~\cite{timnal_exp}.

WTe$_2$ compound was synthesized from elements by reaction of metal with tellurium vapor in the sealed silica ampule. The WTe$_2$ crystals were grown by the two-stage iodine transport~\cite{growth1}, that previously was successfully applied~\cite{growth1,growth2} for growth of other metal chalcogenides like NbS$_2$ and CrNb$_3$S$_6$. The WTe$_2$ composition is verified by energy-dispersive X-ray spectroscopy. The X-ray diffraction (Oxford diffraction Gemini-A, MoK$\alpha$) confirms $Pmn2_1$ orthorhombic single crystal WTe$_2$ with lattice parameters $a=3.4875$~\AA, $b= 6.2672$~\AA, and $c=14.0630$~\AA. We check  that our WTe$_2$ crystals demonstrate large (about 3000\%), non-saturating positive magnetoresistance up to 14~T field, as it has been shown~\cite{ali2014,lvEPL15} for WTe$_2$ and is expected~\cite{jiang15,rhodes15,wang16} for non-magnetic type-II Weyl semimetals~\cite{li2017}.  

We prepare two types of interface structures. One of them is the junction between a gold normal layer and a magnetic Weyl semimetal Ti$_2$MnAl, see Fig.~\ref{sample}  (b).  The other one is the junction~\cite{niwte} between a ferromagnetic nickel layer and a non-magnetic Weyl semimetal WTe$_2$. In both cases, 50~nm thick metallic film (nickel or gold)  is thermally evaporated on the insulating SiO$_2$ substrate. For nickel evaporation, the substrate is mounted on the in-plane magnetized sample holder. 10~$\mu$m wide metallic leads are formed by photolithography and  lift-off technique. Small (about 100~$\mu$m size and 1~$\mu$m thick) WTe$_2$ flakes can be easily obtained from layered WTe$_2$ single crystals. For Ti$_2$MnAl, flakes are obtained by a mechanical cleaving method, see Fig.~\ref{sample}  (a). Then we select the most plane-parallel Ti$_2$MnAl flakes  with clean surface, where no  surface defects could be resolved with optical microscope. A single flake (WTe$_2$ or Ti$_2$MnAl)    is transferred on top of the metallic leads with $\approx 10\times 10~\mu\mbox{m}^2$ overlap and pressed slightly with another oxidized silicon substrate. A special metallic frame allows us to keep the substrates parallel and apply a weak pressure to the sample. No external pressure is needed for a  flake to hold on to a substrate with metallic leads afterward. This procedure provides transparent Ni-WTe$_2$ or Au-Ti$_2$MnAl junctions, stable in different cooling cycles, which has been also demonstrated before~\cite{cdas,nbwte,niwte}.

We investigate transport properties of a single Ni-WTe$_2$ or Au-Ti$_2$MnAl junction by a three-point technique, see Fig.~\ref{sample} (b): a studied contact  is grounded, two other contacts  are employed to apply current $I$ and measure voltage $V$, respectively. To obtain $dV/dI(I)$ characteristics, the dc current $I$ 
  is additionally modulated by a low ac component ($\approx$2~$\mu$A, $f=2$~kHz). We measure both dc ($V$) and ac (which is proportional to $dV/dI$) components of the voltage drop  with a dc voltmeter and a lock-in, respectively.  Measured ac signal is independent of frequency in  1-5~kHz range, which is defined by applied ac filters. In the connection scheme in Fig.~\ref{sample} (b), all  the wire resistances are excluded, which is necessary for low-impedance samples. The measurements are performed in a dilution refrigerator for the temperature interval 30~mK--1.2~K for two different orientations of the magnetic field to the interface. To ensure the homogeneous magnetic state of the junctions,  the magnetization procedure is performed: an  external magnetic field is swept slowly from zero to 5~T, afterward, the external field goes down to zero.

\begin{figure}
\includegraphics[width=\columnwidth]{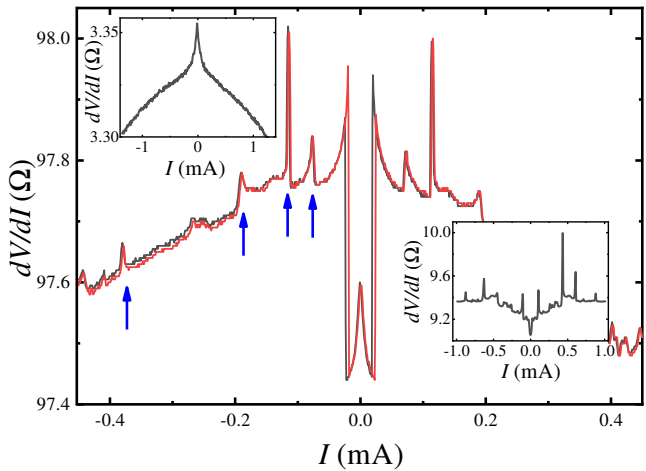}
\caption{(Color online) Typical examples of $dV/dI(I)$ curves  for transport across  Au-Ti$_2$MnAl interface for two opposite current sweep directions. Low-current switchings of $dV/dI$ at $\approx \pm 25 \mu$A bias show well-defined hysteresis. Also,  there are sharp $dV/dI$ peaks at high currents, the peaks' positions are independent of the sweep direction. These $dV/dI$ features  originate from Au-Ti$_2$MnAl interface, since no $dV/dI$ specifics can be observed by four-point measurements for  bulk Ti$_2$MnAl, as depicted in the left inset. The right inset demonstrates similar $dV/dI(I)$ behavior for Ni-WTe$_2$ interface. 
The curves are obtained at 30~mK in zero magnetic field.}
\label{IVs}
\end{figure}

\section{Experimental results}

Fig.~\ref{IVs} provides typical examples of low-temperature $dV/dI(I)$ characteristics for Au-Ti$_2$MnAl  (in the main field) and Ni-WTe$_2$ (in the right inset) junctions. Despite different materials, we observe similar qualitative behavior for both types of the interfaces: $dV/dI(I)$ curves are non-linear, there are   $dV/dI$ peaks at high currents, and sharp symmetric switchings of  differential resistance at low, $\approx \pm 25 \mu$A bias.  The peaks' positions are independent of the current sweep direction, while $dV/dI$ switchings at $\approx \pm 25 \mu$A demonstrates well-defined hysteresis. 

We should connect the observed $dV/dI$ features with interface effects. In a three-point technique, the measured potential $V$ reflects in-series connected resistances of the  Ni-WTe$_2$ or Au-Ti$_2$MnAl interface and some part of the  crystal flake. From $dV/dI(I)$ independence of the particular choice of current and voltage probes in Fig.~\ref{sample} (b), we verify that the  interface resistance dominates in the obtained $dV/dI(I)$ curves. Also, we do not observe any $dV/dI$ features in bulk properties of Ti$_2$MnAl, which is demonstrated by four-point measurements in the left inset to Fig.~\ref{IVs}.  

\begin{figure}
\includegraphics[width=\columnwidth]{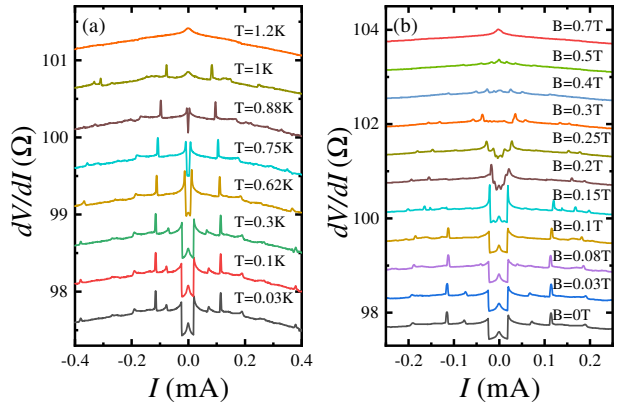}
\caption{(Color online) Evolution of  $dV/dI(I)$ characteristics of the Au-Ti$_2$MnAl junction with temperature (a) and parallel to the interface magnetic field (b). The curves are shifted for clarity. All $dV/dI$ features are suppressed above 1~K or 0.5~T, respectively, they also demonstrate a complicated evolution. The curves in (a) are obtained in zero magnetic field, the ones in (b) are at 30~mK. }
\label{temp}
\end{figure}

The obtained $dV/dI$ features can be suppressed by temperature or magnetic field above 1~K or 0.5~T, respectively, see Fig.~\ref{temp}.  The positions of both the peaks and resistance switchings are moving to zero current with temperature increase until complete disappearance at 1.2~K, as depicted in  Fig.~\ref{temp} (a). 
Evolution of $dV/dI(I)$ curves with magnetic field is different: the width of the low-current region ($\approx 50 \mu$A) is nearly independent of the magnetic field, while the $dV/dI$ switching amplitude is gradually diminishing with the field. In contrast,  $dV/dI$ peaks' positions move to zero in a complicated manner. 

The detailed behavior of  $dV/dI$ peaks' positions is shown in Fig.~\ref{magn} for Au-Ti$_2$MnAl junction for parallel (a) and normal (b) to the interface magnetic fields. For both field orientations, the positions of the peaks are shifting non-monotonously to smaller currents, so the peaks disappears above some value of magnetic field. This value is significantly smaller for the normal field orientation ($\approx 0.2$~T, see Fig.~\ref{magn} (b)), in comparison with $\approx 0.6$~T for the parallel one (a).

To our surprise, not only $dV/dI(I)$ curves are similar for Ni-WTe$_2$ and Au-Ti$_2$MnAl interfaces in Fig.~\ref{IVs}, but also $dV/dI$ features show analogous behavior.  For  Ni-WTe$_2$,   $dV/dI$ peaks' positions are shifting  to zero current with magnetic field, the suppression is twice faster in normal field, see Fig.~\ref{magn1}.

\section{Discussion} \label{disc}

\begin{center}
\begin{figure}
\includegraphics[width=\columnwidth]{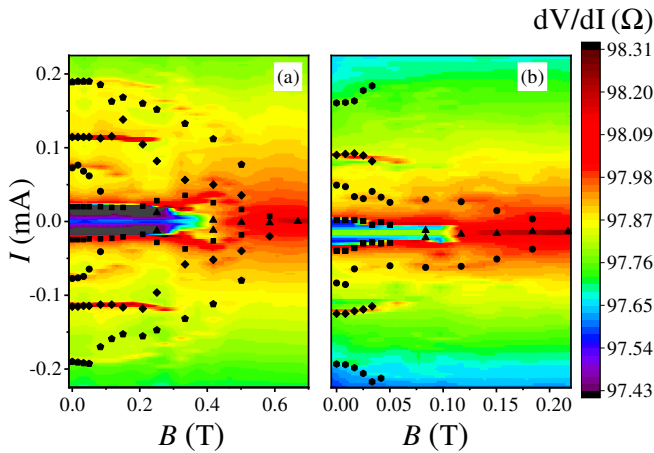}
\caption{ (Color online) Evolution of  $dV/dI$ peaks' positions for Au-Ti$_2$MnAl junction for parallel (a) and normal (b) to the interface magnetic fields. $dV/dI$ peaks are shifting  to lower currents with the field increase. Full peaks' suppression can be seen at $\approx 0.6$~T for the parallel field orientation, but it occurs much earlier, at  $\approx 0.2$~T, for the normal one. The data are obtained for 30~mK temperature.
}
\label{magn}
\end{figure}
\end{center}

\begin{center}
\begin{figure}
\includegraphics[width=\columnwidth]{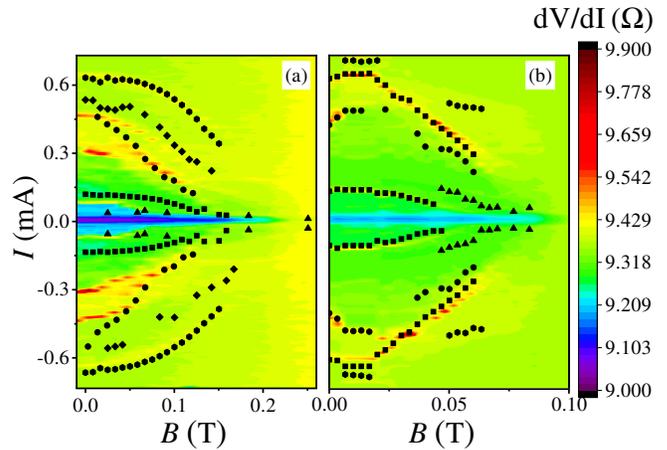}
\caption{ (Color online) Evolution of  $dV/dI$ peaks' positions for Ni-WTe$_2$ junction also for parallel (a) and normal (b) to the interface magnetic fields. The behavior is qualitatively similar to the Au-Ti$_2$MnAl case in Fig.~\ref{magn}. $dV/dI$ peaks are suppressed at $\approx 0.2$~T for the  parallel field orientation, the suppression is twice faster in normal fields.   The data are obtained for 30~mK temperature.
}
\label{magn1}
\end{figure}
\end{center}

As a result, both Au-Ti$_2$MnAl and Ni-WTe$_2$ junctions demonstrate similar $dV/dI(I)$ characteristics, with hysteresis at low currents and sharp peaks  at high ones. Moreover, we observe qualitatively similar evolution of the peaks' positions with magnetic field  for both structures in Figs.~\ref{magn} and~\ref{magn1}. For this reason, the obtained results should have the same origin for these structures. From the experimental point of view, the obtained $dV/dI(I)$ curves are similar to ones for ferromagnetic multilayers~\cite{myers,tsoi1,tsoi2,katine,single,balkashin,balashov}.

\begin{center}
\begin{figure}
\includegraphics[width=\columnwidth]{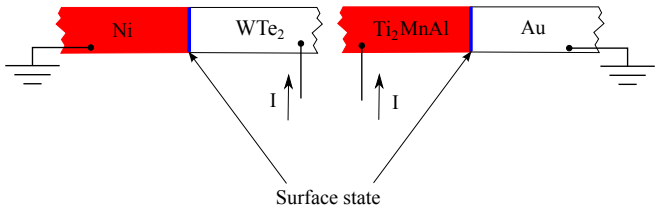}
\caption{ (Color online) Schematic representation of Au-Ti$_2$MnAl and Ni-WTe$_2$ interfaces, which are characterized by Weyl surface states at the interface (blue color). In every case, one side of the junction has significant net spin polarization of carriers (Ni or Ti$_2$MnAl, respectively, red color). Thus, spin-dependent transport through the Weyl surface state is investigated for these junctions.
}
\label{discussion}
\end{figure}
\end{center}

Weyl surface state is the only common characteristic~\cite{armitage,das16,feng2016,timnal} of Au-Ti$_2$MnAl and Ni-WTe$_2$ interfaces, since  the materials are completely  different for the metallic contacts and the semimetals itself: for Au-Ti$_2$MnAl junction we study transport between  a magnetically-ordered WSM and a normal metal, while Ni-WTe$_2$ one represents the junction  between a non-magnetic WSM and a ferromagnet, see Fig.~\ref{discussion}. Also, strong temperature dependence in the  30~mK-1.2~K range  can only originate from the surface, since transport properties of Ni or Au layers and  bulk WSM~\cite{lvEPL15,timnal_exp} are invariant in this temperature range.

For Au-Ti$_2$MnAl and Ni-WTe$_2$ samples,  one side of the junction has significant net spin polarization of carriers (Ni or Ti$_2$MnAl, respectively). We should conclude, that similar $dV/dI(I)$ curves are produced by spin-polarized transport through the Weyl surface state at the interface. In some sense, our experiment resembles ones on ferromagnetic multilayers,  where spin-dependent scattering affects the magnetic moments of the spin-polarized layers, while their mutual orientation defines the differential resistance~\cite{myers,tsoi1,tsoi2,katine,single,balkashin,balashov}. It might be natural~\cite{topinssurf,current}, that we observe similar  $dV/dI(I)$ characteristics.  

Let us start from $dV/dI$ switchings at low currents in Fig.~\ref{IVs}. At zero bias, one can expect that spin polarization of some carriers at the WSM surface  is aligned parallel to one in the ferromagnet due to the complicated spin texture~\cite{texture,araki} on the Weyl surface~\cite{jiang15,rhodes15,das16,feng2016}.
For this reason, even spin-polarized carriers have a direct transport channel, which is reflected in low junction resistance at zero bias.  While increasing the current through the junction, spin-momentum locking produces~\cite{current,araki} a preferable spin polarization in the surface state, which is reflected as sharp  $dV/dI$ increase for both signs of the current. As usual~\cite{myers}, current-induced switchings are accompanied by hysteresis in  Fig.~\ref{IVs}.  Spin alignment   disappears at zero bias, when high magnetic field or temperature destroys the spin textures in the topological surface state, see Fig.~\ref{temp} (a) and (b).

Similarly to the ferromagnetic multilayers~\cite{tsoi1,tsoi2}, we should identify $dV/dI$ peaks in Fig.~\ref{IVs} as the onset of the current-driven magnon excitations. However,  evolution  of the peaks' positions with magnetic field is unusual: the peaks  are moving to lower currents in Figs.~\ref{magn} and~\ref{magn1},
which is opposite to the known bulk magnon behavior~\cite{myers,tsoi1,tsoi2,katine,single,balkashin,balashov,cosns}. This peaks' evolution is the main difference of our results from the standard magnon experiments~\cite{myers,tsoi1,tsoi2,katine,single,balkashin,balashov,cosns}.

Since the peaks disappear simultaneously with $dV/dI$ switchings in Fig.~\ref{temp} (a) and (b), we should also connect~\cite{topinssurf} the magnon excitation with  spin textures~\cite{texture,araki} in the topological surface states.  In general, the $dV/dI$ peak  position $I_{sw}$ is described by Slonczewski model~\cite{slonczewski,katine}. Slightly simplified,
\begin{equation}
I_{sw}(H) \sim \alpha \gamma e \sigma H, \label{eq}
\end{equation}
where $\alpha$ is the damping parameter, $\gamma$ is the gyromagnetic ratio, $\sigma$ is the total spin of the free layer. In contrast to multilayers, the total spin 
$\sigma$ is not a constant. It is diminishing to zero when  high magnetic field or temperature destroys the spin textures in the topological surface state. This $\sigma (H,T)$ dependence  can be the origin of the unusual peaks' evolution in Figs.~\ref{temp},~\ref{magn} and~\ref{magn1}. However, we have no complete description of the magnon dynamics in Weyl topological surface states, in contrast to the case of topological insulators~\cite{topinssurf}.

It is well known, that surface state transport can be observed only at temperatures, which are significantly lower than the temperature of the corresponding spectrum gap. For example, for the quantum Hall effect, temperatures below 1~K were necessary to see the edge state transport~\cite{muller,alida,MZ}. The same
situation is for topological insulators~\cite{amit,kononov}. For Weyl semimetals, we also observed~\cite{nbwte} Weyl specifics in Andreev reflection only
below 1 K, while the Nb gap was estimated as about 10~K in this experiment. Strong temperature dependence in Fig.~\ref{temp} (a) is in a contrast with known bulk behavior~\cite{myers,tsoi1,tsoi2,katine,single,balkashin,balashov,cosns}, which can also indicate the surface state origin of the observed peaks.

\section{Conclusion}

As a conclusion, we experimentally  compare two types of interface structures with magnetic and non-magnetic Weyl semimetals. They are the junctions between a gold normal layer and magnetic Weyl semimetal Ti$_2$MnAl, and a ferromagnetic nickel layer and non-magnetic Weyl semimetal WTe$_2$, respectively. Due to the ferromagnetic side of the junction, we investigate spin-polarized transport through the Weyl semimetal surface. For both  structures, we demonstrate similar current-voltage characteristics, with hysteresis at low currents and sharp peaks in differential resistance at high ones. Despite this behavior resembles the known current-induced magnetization dynamics in ferromagnetic structures, evolution  of the resistance peaks with magnetic field is unusual.  We connect the observed effects  with current-induced spin dynamics in Weyl topological surface states.

\acknowledgments

We wish to thank  V.T.~Dolgopolov for fruitful discussions, S.S~Khasanov for X-ray sample characterization, A.A.~Kononov and O.O. Shvetsov  for help with experiment.  We gratefully acknowledge financial support partially by the RFBR  (project No.~19-02-00203), RAS, and RF State task.

\end{document}